# The Integration of Blockchain and Artificial Intelligence for Secure Healthcare Systems

Dr. Umar Safdar
Center of Science and Engineering
New York University
Umar.safdar@nyu.edu

Dr. Simon Gabrael
Center of Science and Engineering
New York University
Simon.Garbael@nyu.edu

## Abstract

Verisign reported a 125 percent increase in data breaches within the healthcare sector in the United States during 2022, with 18.2 million patient records being impacted. Growing healthcare data volumes and diversification mean that medical information is becoming more valuable. Many Health Centers use various technologies to ease the classification, storage, and exchange of big data. This use can also make the health data of the users at risk and vulnerable. AI and blockchain are among the leading technologies at hand. With AI, data-driven operations and big data efficiency have been improved with respect to traditional techniques. Due to its potential to bring about improvements in health services and lower medical costs, this AI technology is regularly used in healthcare. Blockchain helps protect transactions on sharing information and private privacy as long as the exchange of knowledge is that of the standard.

The objective of this analysis is to investigate the research and unique contributions since 2008 regarding blockchain-integrated AI and healthcare systems. The work sheds light on applied AI-based healthcare schemes with machine, ballistic, and acrylic learning and disparate blockchain structures. The use of technology in order to ensure patient data security and manage medical information effectively in healthcare settings offers a highly successful position for both healthcare providers and patients. From 2018 to 2021, the best year was 2021 to grow, enhancing everything to examine the download of the device and the counting of Google Academies, for which the joining perspective was borrowed; local research experts were asked, identified articles in recent years, and read reviews of large research grants.

# 1. Introduction

Contemporary healthcare delivery is complicated, dependent on several aspects, and impacted by a variety of statements. Healthcare is no longer the domain of individual healthcare professionals or healthcare providers, but rather an extensive, multipartite arrangement across multiple countries. Additionally, it necessitates the convergence of various groundwork areas, including cutting-edge technology, engineering, and biomedical science, an integral part of multidisciplinary studies as a collective formation. In such a structure, data, bi, tri, and quadraphasic data security and health data confidentiality are of paramount health importance, demanding ongoing research and growth. Novel advances in cybersecurity, such as blockchain properties, can be advantageous in the modernisation and reform of healthcare services. One way to interpret in simplified concepts is unique slightly extended unforfeitable electronic notepad for storage of data in collection strings, entitled blocks [1]. Each additional string, archiving tens of thousands of statements can be combined into a block, serving as distinct electronic descriptions or orientations of that collection that may be made accessible for a discrete period. AI is progressing exponentially and is assumed to hold the possibility of overtaking different dispatches [2]. Major innovative tech firms are currently active in AI, as are original cyber attackers, as it can be utilized for security and health challenges.

## 1.1. Background and Significance

1.1.1. Background The advancement of technology in the fields of healthcare, artificial intelligence (AI), internet of things (IoT), and blockchain has significantly influenced new methods to provide patient care. The development of telemedicine is one of the mainstream implementations in healthcare fields. This enables patients and healthcare providers to interact through smartphone applications or websites. Doctor diagnoses on the analysis of the medical report [3]. However, a doctor without advanced tools may give their subjective evaluation of the diagnosis of the medical images. The unstructured data or patient health records become vulnerable and easy to alter. This paper provides the solution based on the improvement of the existing system.

The healthcare domain involves an enormous amount of data to be stored and shared. As the need for healthcare applications increases due to chronic diseases and other medical emergencies, the amount of data being shared is growing exponentially [2]. Currently, healthcare systems are using traditional approaches to store and manage data and are often centralized databases. Many hospitals and healthcare organizations do not sufficiently secure their system and possess privacy issues [4]. According to the HIPAA act, patient privacy must be maintained; however, due to data breaching or unauthorized data sharing, there are many issues in healthcare systems [1]. There have been many studies exploring some possible improvement strategies in healthcare systems. Still, there is a lack of clear proof of concept (POC) to show the implementation of blockchain and artificial intelligence technology significantly increases security and operational efficiency.

1.1.2. Significance The implementation of blockchain and artificial intelligence (AI) for secure healthcare applications is highlighted in this paper. Clear proof of concept (POC) incorporates blockchain and AI technologies and demonstrates a development system. Specifically, this research significantly focuses on the security enhancement of data that are shared in the healthcare system and records patient discharge summary [5]. A conceptual framework is proposed by implementing the transfer learning concepts. In this concept, the research recognizes a language model in the medical field as source data that is shared among healthcare providers collaboratively to improve medical fields. As a response, the recommended medication field is enhanced in the medical records. Furthermore, the research also examines and discusses the scalability and applicability of AI technology with the blockchain mechanism to secure patient records, which supports machine learning modeling to adapt and improve the medical fields over time.

A comprehensive examination of blockchain technology is conducted by using Kamran Ayub's SMART Incubator health model, which provides a mechanism to apply this protocol [6], and it is discovered that amalgamation with AI technologies enhances transparent healthcare and efficiently secures the medical data record. The reviews of existing work are more focused on the formation of patient data security and decentralized health services. Finally, an additional possible exploitation possibility includes a secure medical image sharing framework, technology proposed by a secure inner source, and a AI technology concept of medically explainable emotional recognition with digital stethoscope.

## 2. Blockchain Technology in Healthcare

The blockchain: safeguarding and sharing health data securely - As each block in a blockchain contains a pointer of the previous block, any data violation on one block can always be detected from even a single data point. As data becomes more decentralized, the blockchain significantly hinders the hacker's ability to breach data integrity. Blockchain technology, as a digital ledger, exhibits great potential in maintaining the integrity and security of health-related information, and is an epoch in secure and scalable e-Health implementation [7]. A blockchain is a database that can only be added to and not modified or removed from previously entered records. Each block in a blockchain contains the address of the previous block, making the data highly traceable, safeguarding data from potential tampering. The most popular form is the public blockchain with an open network that allows access to all. In comparison, the private blockchain can only be accessed by specified participants for the network [1]. Blocks in a blockchain are formed from transactions or agreements, and once written, can never be altered. However, from the refreshment and reemployment of these definitions in contemporary political theory blockchain can clearly be seen as a set of tools and processes that possess the powers of transferring, sorting, and storing data in specific manners. Blockchain rapidly emerged in health policy discourses and practices for its abilities to streamline opaque bureaucratic processes, authenticate the veracity of transactions, and therefore beyond doubt be reinstitutionalized [8].

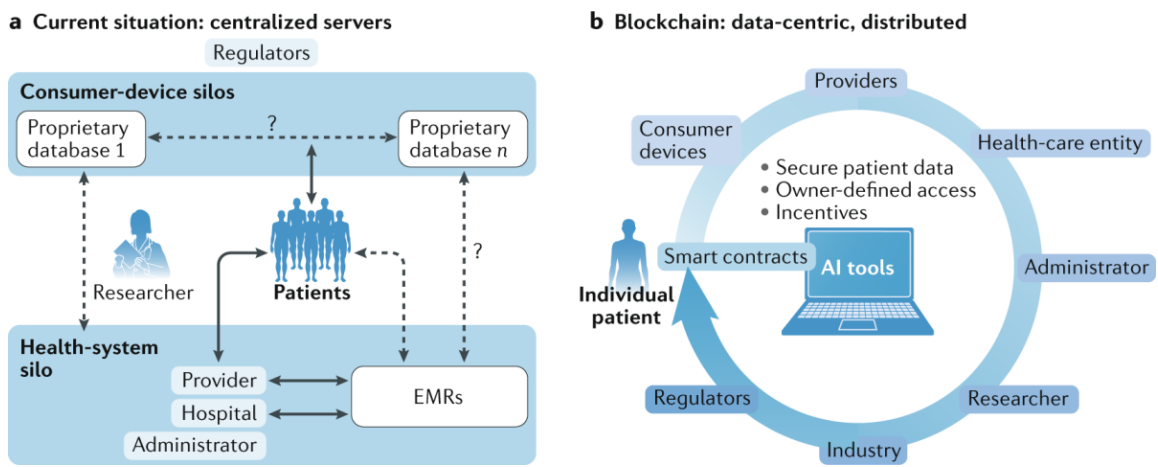

Fig1. Use of Blockchain & AI Operational lifecycle

More convincingly than the accepted possibility of decentralized autonomous and granulated governance (known as 'smart contracts' run in blockchain-based platforms, solidification is the technology's contemporary destinations and purposes. Because of benefits like traceability and transparency, blockchain is attracting attention in pharmaceutical and health data. Four primary health applications of the blockchain that may be driven are (i) electronic medical records management, (ii) prescription drug traceability, (iii) peer-to-peer data settlement, and (iv) patient authorisation controls. The established Healthureum application, which is scheduled to be introduced in 2020, will rely on the managerial and technical expertise of the health and blockchain industries to establish an extensive ecosystem that will permit troublesome administrative issues commonly linking health services and insurance companies to be passed, underpinned by a future HSX token. For example the Drug Supply Chain Security Act — DSCSA — that requires pharmaceutical supply chain stakeholder engagement. At every phase in the supply chain, manufacturers, repackers, wholesale distributors, dispensers, and third-party logistics providers, at a minimum, must record product transfers. Because such parties are required to document only who they transacted with, the estimated economic value of the DSCSA has been significantly revised downward, with an estimated net cost to be min. $2.77 billion from 2017-2027.

## 2.1. Key Concepts and Principles

Blockchain technology, with its novel and solid characteristics such as decentralization, transparency, resistance to tampering, secure ledgers of transaction information, and automation through smart contracts, is considered a breakthrough for healthcare [2] [1]. Research has been conducted on security improvement mechanisms in various AI-based healthcare systems through blockchain technology, and analysis of the implementation status and corresponding research trends regarding this in the healthcare system. A case study based on blockchain technology for security enhancement in lenses AI (Artificial Intelligence)-based livestock diagnostic support services has been conducted to understand the practical feedback of such technology implementations. Blockchain technology application's potential in improving the security of various AI-based healthcare systems and relevant issues have been identified and discussed.

The concept of blockchain was initially presented in 2008 together with Bitcoin. After that, both public and private sectors have focused on blockchain technology due to its characteristics: decentralized, transparent, and tamper-resistant. Blockchain has been widely applied to various developed services, supply chains, logistics, erasure management, finance, and even public institutions, based on the internet environment [9]. As it provides the advantage of unalterable ledger information management by connecting data with specific hash algorithms, many sub-concepts and applications along with their terms have been introduced into the technological field: blocks, chains, nodes, and ledgers. Pointing to chain-based data recording structure abstracted as a block, blockchain is a special technology scheme that stores all participating nodes and their transactions as a decentralized and controlled network. By utilizing a cryptographic hash function, blocks are encrypted and saved as a part of a chain network. Consensus mechanisms like Proof of Work

(PoW) and Proof of Stake (PoS) support the system maintaining each transaction validation, which represents how nodes agree on what the system state would be. Regarding security issues in terms of Peer-to-Peer (P2P) services and cryptographic user anonymity, blockchain can be a suitable scheme for using a specific data stratum by applying a hash function on it [10].

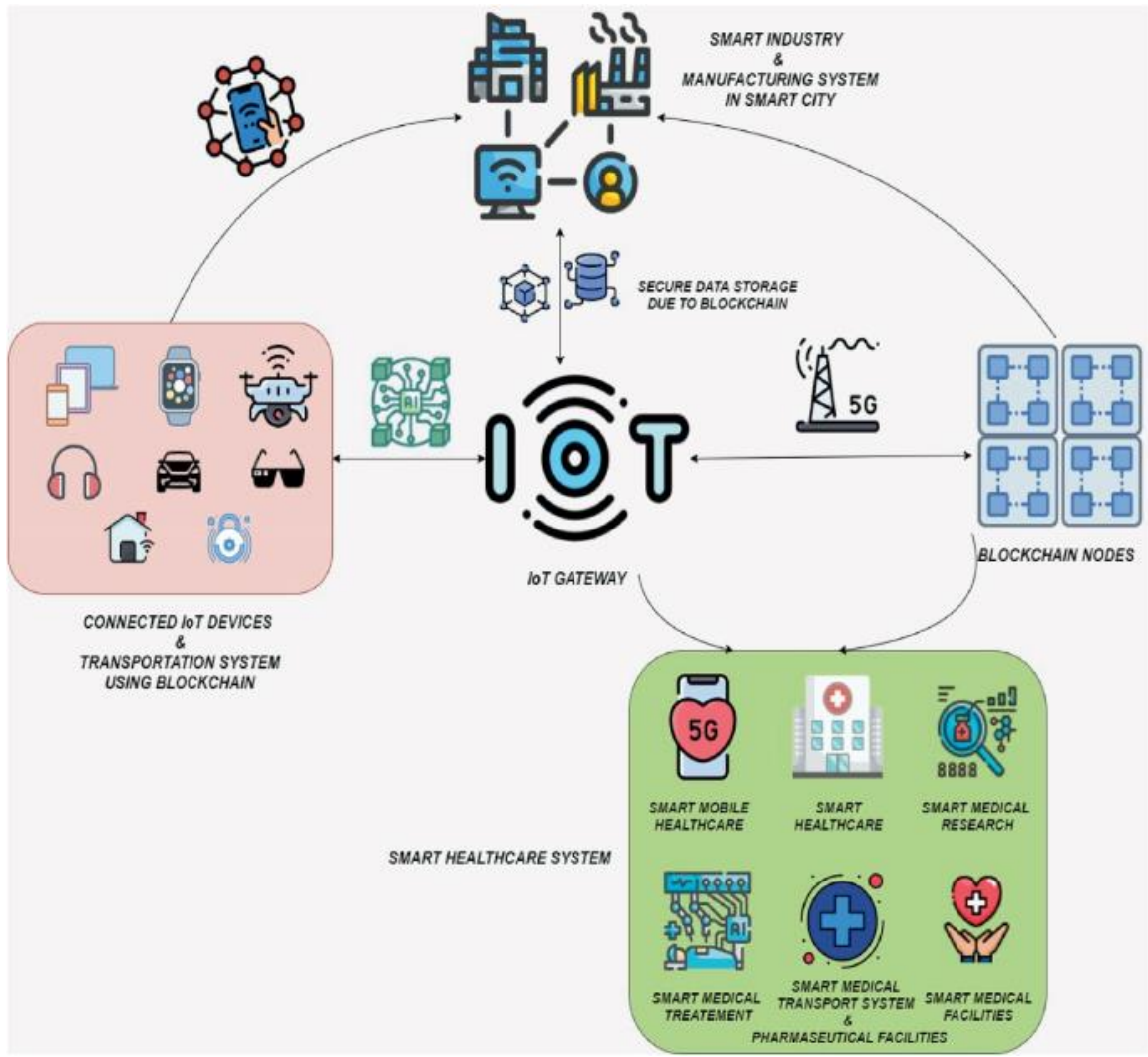

Fig. 2. SMART Health with Blockchain and AI

## 3. Artificial Intelligence in Healthcare

One of the first-generation cryptocurrencies 'Bitcoin' and the underlying technology 'Blockchain' has drawn extensive research attention in the last decade, supporting the potential adoption in various domains beyond financial services such as gold storage, cloud computing, estate, and banking. Medical and health care areas are the foremost to embrace its benefits, offering the potential to redesign healthcare data storage and transactions over medical records, therefore realizing an inclusive, fair, protected, and high-quality clinical environment. In developing countries such as Korea with a high reliance on some main superintendents, healthcare facilities, and well-known clinical data commercial firms for medical data, de-identification methods and tightly connected patents are regularly practiced for concealing highly classified whereabouts and therapies of medical data. However, intermediaries and managers often carelessly or maliciously disburse algorithms or information, exposing vulnerabilities to ways. Blockchain technology effectively handles privacy and security vulnerabilities associated with medical record data in medicine and public health through methods based on pairwise encryption. Common adverse drug effects can be quickly shared and rapidly intervened in daily clinical trials with similar strategies and manipulation of medical records. Given the greater risk trading of chronic diseases in hospitals, preferred doctor's offices and employees' services could be granted an acceptable and manageable approach for exploring common interests and balancing the partnership amount in the absence of pre-assessed legalities [11]. Safe and open intellectual contracts significantly prevent potential risks such as privacy misuse, short messages, and legal questions through rule-based evidence. This technology has been shown to have the shortest therapeutic duration and the fewest re-hospitalizations compared to already established procedures for depression patients.

### 3.1. Applications and Benefits

**Applications and Benefits**

Artificial intelligence (AI) is thriving, particularly in healthcare, giving rise to improved methods for diagnostics, tailored therapy, operational efficiencies, and other applications. Such achievements are made through the benefits contributed by analytical algorithms and immense learning methods among the AI loom. There is a substantial increase in developing and applying predictive devices and diagnosis software that are driven by AI [12]. The benefits include enhancing the ability to distinguish anomalies (for instance, X-rays and CT scans), which aids in ensuring more precise, tailored therapy options. Moreover, due to its potential to engage with sophisticated datasets, AI included in the analysis sector can anticipate diagnoses and other often unattainable data.

To classify, applications of AI that are growing and their benefits to healthcare are indicating herein. Diagnosis is foremost. Although stakeholders' understanding of AI algorithms is still growing, there are already app examples comparing the patient's health data (such as a blood test, heart rate pattern, and eye check) to medical data collated from various sources. Such complex datasets are securely stored and analyzed by an AI algorithm running on a cloud-based evaluation platform, delivering examination accurately. Concerns about AI-enabled health tools and products and how they are implemented would need to be addressed. At the same time, the stakes about the security and reliability of analysis carried out by the non-human entity must be adequately addressed [7].

Using the prediction, AI algorithms can also propose a treatment plan. The app can have access to a wide range of data concerning a person's health circumstances and offer a tailored remedy plan. However, it is indispensable to acknowledge that AI algorithms are designed to provide a recommendation and their suggestion might not be the only strategy achievable. AI devices have recently appeared in research and therapy to generate a tailored therapy plan. These devices pull together assorted health information about a person and employ AI to offer several options for therapy strategy. Another emerging application of AI in healthcare is to boost the effort. This can be done by using AI devices in virtual health aide apps that give information relating to health or first aid. There are also telemedicine technologies that provide impartialized consultation based on health data analysis by AI [13]. Interestingly, there are AI uses to forecast patient-specific disease and staff needs. As such, the solution plan falls within the remit of the eHealth trust framework [2].

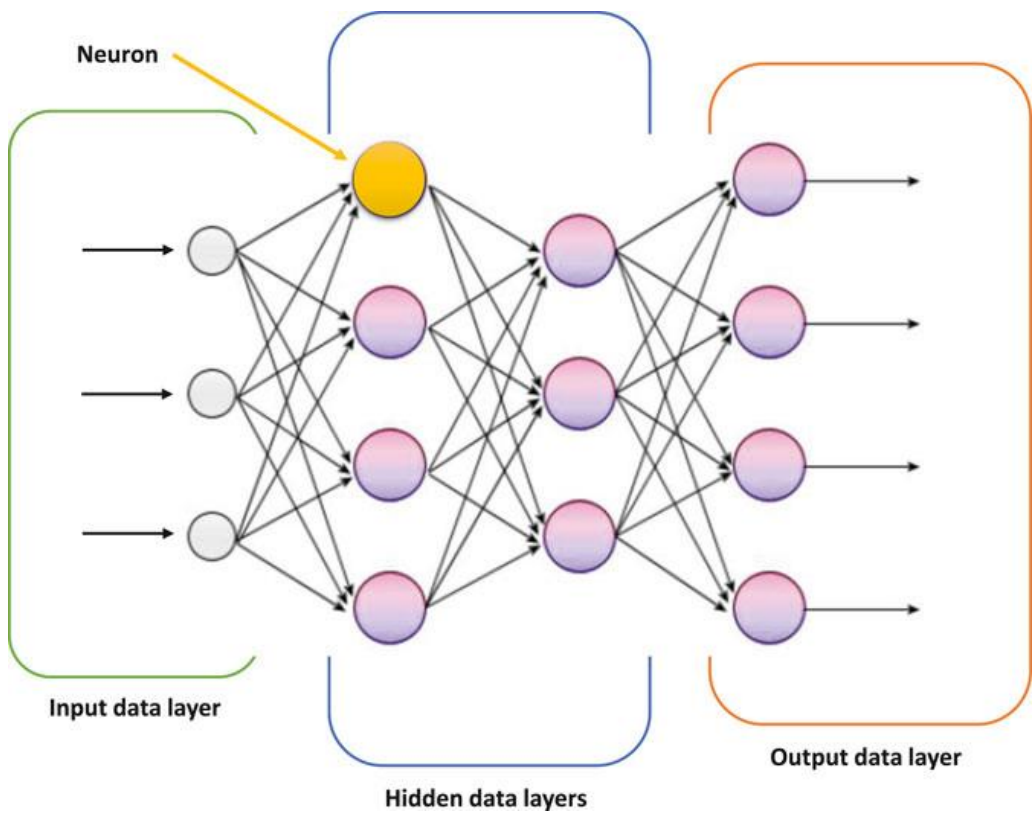

Fig. 3. Different layers in Blockchain AI with the interaction model

## 4. Integration of Blockchain and AI in Healthcare

Safety is one of the most crucial components in the healthcare industry. Because of having safety issues, patient's demographics, contact details, and other medical information can't be shared with numerous personnel. To quicken and automate the decision-making process in healthcare systems, a model then has been proposed which combines AI and blockchain. The current medical history is collected by AI [14]. It also analyses the data to examine the patient's present condition and make a decision. In hospitals or clinics, AI has a massive database of a disease that can be used to forecast the predicament. The data is later saved in medical chains, which is completely safe. With new conditions, training accuracy also boosts since the database is enhanced.

The proposed approach allows the computer to work automatically and independently, rather than relying on time-consuming clinical evaluations. There are numerous patients and numerous doctors in huge hospitals. Essential data (demographics, past diseases, current condition, precise diagnosis, etc.) are stored on a public chain instead of on computers at different hospitals, allowing any physician to obtain them. Scaling up healthcare systems is a remarkably intricate task. There isn't one role for the entire medical team since each individual has assorted tasks. Computer systems are various in different health organizations. The AI with blockchain approach is to track all processes involving a patient's data as an endorsement from all members of the network.

As a result, if there is any dishonesty or inauthenticity in a patient's data, it is quite easy to trace back. The entire treatment process (from consultation to procedures) will be stored and may be analyzed using the AI algorithm to refine the decision-making system. This combination of blockchains from AI has opened the door to distribute nodes. The blockchain combined with AI can boost healthcare safety systems by detecting any abnormal activity and maybe taking predeterminative movements.

### 4.1. Use Cases and Examples

The opportunity for interlinking blockchain and artificial intelligence (AI) technologies for healthcare setups is broad-ranging. A sweeping examination transpires through this configured approach including real-world illustrations embodied by remarkable implementations, pilot projects, and clinical studies. Additionally, barriers to progress, viewpoint points, and inherent barriers are conferred emphasizing implementability at the same time making it farfetched to obtain diverse use cases that consolidate AI and blockchain technologies in healthcare endeavors. Nonetheless, the established use cases occur in both scalable and clinically conceivable ways garnering brand postulates for this configuration.

The realm of healthcare yields infinite break points for mistakes as indicated by the World Health Organization. A predominantly restrictive hallmark of intense situations with conflicting care suppliers is the swift yet well thought regarding procedure each case calls for a favored treatment paradigm. Notwithstanding faith and ecological elements influence the ultimate choices elaborated by care suppliers, manifold appropriate treatment arrangements might exist. In order to avert human blunder, the procedure regarding treatment best taken is enforced via AI [15]. Nonetheless, this part implicates the aggregation from numerous sources of care as well as ultimate care rendered. This can be formidable as incorrect inputs endanger the encompassing remedy. In the setup of healthcare in particular, this can engender illicit behavior with major consequences. Each such resolution shan't be based on the true better good for the patient. The 2nd hallmark of this configuration is the block on treatment decisions via the patient contingent on a secure educated guess with respect thereto. [7] By entrusting the patient's tolerant consent to the blockchain this generates a robust track timesheet thereto patient piece of advice which isn't susceptible to modification and all actors involved can proceed from. Substantiating the consent is positioned via facial recognition on an immobile side is based on the very first reading of patient consent being virtual in nature. Submissive an inspiring effort was conducted in the deployment of this desired functionality within an extreme but potentially difficult adoption venue such as intensified care units (ICUs). Disregarding the crystallized results exhibiting a 75% decrease of command and record error on the compassionate and a 25% lowering of regimen-associated mistake, a broader conversation with respect thereto decentralized technology enforcing multi-benefits, while simultaneously heeding to the expectation that overly zealous expectations from this implementation probably are difficult to fulfill.

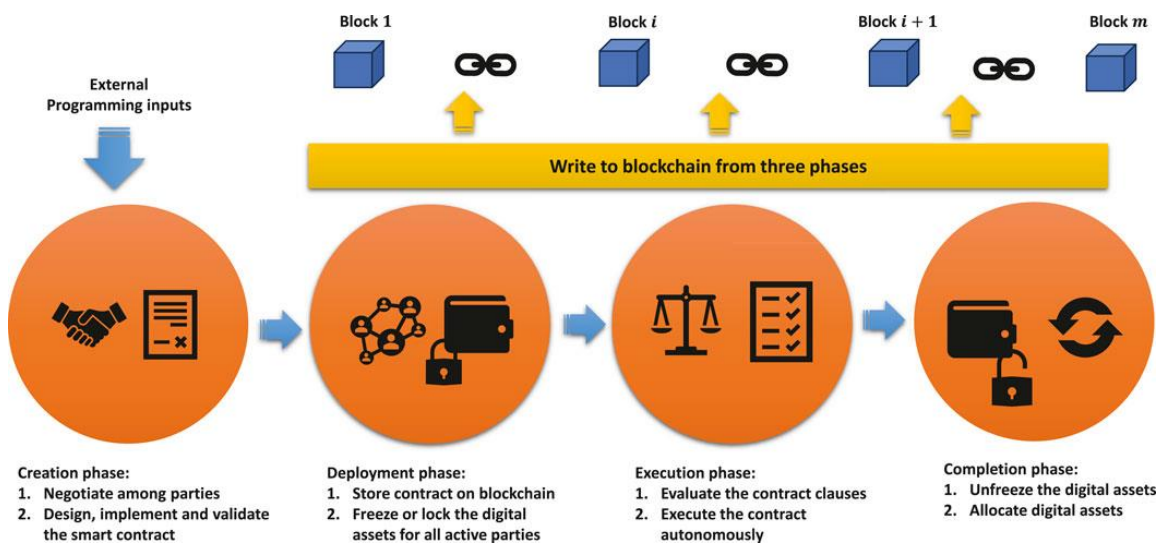

Fig. 4. Different phases of Interactions in Blockchain with AI

## 5. Challenges and Opportunities

Amid an increasing array of blockchain applications, the healthcare industry will be revolutionized by the integration of blockchain technology and artificial intelligence. Far more transparency in the healthcare industry is provided by blockchain technology applications. It improves data sharing capabilities and opens up the potential for information that is otherwise critical for the establishment of health care strategies. Integrating AI will enhance blockchain innovations in healthcare, making interoperability transparent and simple. Both healthcare regulators need to work together to approve and incorporate these emerging technologies and foster the establishment of security installations that are smarter and more secure. Implementation of advanced technology may initially not be that simple because of the current systems' complexity and audit requirements, and this might create a delay in acquiring and compounding data even further concerning emerging technology in healthcare [2]. A possible downside is that healthcare professionals and institutes could choose to disregard the initiative. Because they may not be willing to adjust their management model, there might be a barrier to modification. Because healthcare systems differ according to their custom software and systems, it might evolve into a daunting task to adapt them into the AI model. Healthcare professionals would have to squander valuable time attempting to enlighten developers concerning their practises. This is not economically viable, considering the existing overwhelming efforts. This would then take significantly longer for AI and Blockchain technology to be adopted widely in the healthcare sector, defeating its purpose for improving the effectiveness and safety of the healthcare system.

### 5.1. Ethical and Legal Implications

This article highlights both the benefits and limitations, including ethical and legal concerns, addressing the use of these technologies in a healthcare context. There is a growing, and often enthusiastic, interest in healthcare potential applications of blockchain technology and artificial intelligence. Despite numerous claims in favour of, and against, the prospective positive impacts, a balanced view suggests both considerable potential benefits and risks. For widespread adoption, it is essential to address these concerns in a balanced and informed manner.

There is a growing recognition of numerous potential applications across all four broad categories – clinical management, financial management, public health, and research. The most commonly cited applications among the reviewed articles are EHR management, payment and claim management, drug traceability or inventory management, and sensors management (i.e., IoT, temperature, humidity, and humidity sensors), advertisers, and alarms. Interestingly, these applications comprise a high percentage of fraud research [16]. Additionally, some early pilots with successful prototypes have encouraged speculation on potentially transformative impacts. On top of this, more commercial interests are beginning to emerge, with an increasing number of providers marketing solutions. All of this has combined to generate considerable excitement among regulators and policymakers.

However, a closer look also highlights numerous concerns and challenges – many of which are rarely acknowledged, let alone addressed. Most pressing is the lack of systematic robust evidence that these technologies can indeed deliver on the claims. A review found that current studies are still in the exploratory and descriptive stages, underlining the need for empirical evidence of the actual benefits brought by these technologies in healthcare applications, as well as the shortcomings and limits once their adoption begins to spread more extensively. This is a critical gap given the potential risks.